\documentclass[preprint,12pt,3p]{elsarticle}

\usepackage{amssymb}
\usepackage{multirow}

\usepackage{doi}

\usepackage{caption} 
\usepackage{graphicx,subfigure}
\usepackage{verbatim}
\usepackage{makecell}
\usepackage{longtable}

\usepackage{rotating}

\usepackage{lineno}

\journal{Nucl. Instrum. Methods. Phys. Res. A}

\begin{document}

\begin{frontmatter}

\title{ A reactor antineutrino detector based on hexagonal scintillator bars }

\author[label1,label2]{Mustafa Kandemir}

\address[label1]{Department of Physics Eng., Istanbul Technical University, 34469, Istanbul, Turkey}
\address[label2]{Department of Physics, Recep Tayyip Erdogan University, 53100, Rize, Turkey\fnref{label4}}

\cortext[cor1]{corresponding author}

\ead{mustafa.kandemir@erdogan.edu.tr}


\author[label1]{Altan Cakir\corref{cor1}}
\ead{cakir@cern.ch}


\begin{abstract}

 This study presents a new concept of segmented antineutrino detector based on hexagonal plastic scintillator bars for detecting antineutrinos from a nuclear reactor core. The choice of hexagonal scintillator bars is original and provides compactness. The proposed detector detects antineutrinos via inverse beta decay (IBD) with the prompt-delayed double coincidence. Owing to its segmented structure, the background, which satisfies the delayed coincidence condition can be eliminated by applying proper event selection cuts. In this manner, the main focus is to determine proper selection criteria to precisely tag the true IBD events. Monte-Carlo simulation is carried out to understand the characteristic of the IBD interaction in the proposed detector by using Geant4 toolkit. A set of event selection criteria is established based on the simulated data. It is found that a detection efficiency of 10$\%$ can be achieved with the selection condition applied. It is also shown that fast neutrons, which constitute the main background source for above-ground detection, can be effectively eliminated with these selection criteria. The motivation for this study is to install this compact detector at a short distance ($<$100 m) from the Akkuyu Nuclear Power Plant, which is expected to start operation in 2023.
    
\end{abstract}

\begin{keyword}
 Antineutrino detector \sep Reactor monitoring \sep Hexagonal plastic scintillator bar  \sep GEANT4
\end{keyword}

\end{frontmatter}


\section{Introduction}
\label{sec1}

Akkuyu Nuclear Power Plant (NPP) is under development in Akkuyu, in Mersin province, Turkey. It will be the country's first nuclear power plant. The reactor will use Rosatom's third generation VVER-1200 design, and it is expected to enter into operation in 2023. The reactor will be composed of 4-Unit and each unit will have a power of 1200 MWe. In addition to energy production, Akkuyu NPP will provide the opportunity to test neutrino oscillation studies and enable to study neutrino physics applications.

Nuclear reactors are the powerfull source of antineutrinos. In a typical power reactor, the contribution to the reactor thermal power mainly comes from the fission of four main fuel isotopes: U-235, Pu-239, U-238, and Pu-241. The fission products formed as a result of fission of these fuel isotopes are mainly neutron-rich nuclei and therefore undergo beta decay to become stable nuclei and emit antineutrinos. Since each fission process relases about 200 MeV of thermal energy and $\sim$6 antineutrinos, a total of $\sim10^{25}$ antineutrinos are emitted per day isotropically from a 1 GW power. Although antineutrinos interact with the matter very weakly, the high antineutrino flux from the reactors allows to detect antineutrinos even with a compact detector installed close to the reactor.  

In a light water reactor (LWR), as the reactor fuel burns up, the contribution of fuel isotopes to fission varies over time. Since the emitted detectable neutrino spectrum per fission of each isotope differs, both the number and energy of the emitted neutrinos change over the course of the fuel cycle. The change in fissile content or more clearly the "burn-up effect" can be revealed by measuring antineutrinos. As a matter of fact, this phenomenon was first demonstrated at Rovno \cite{Rovno} nuclear reactor in Russia, and then a few years later in modern SONGS \cite{Sandia1,Sandia2,Sandia3} experiments.

The relation between the antineutrino detection rate, the thermal power, and the fuel burnup can be expressed as \cite{Bernstein}

\begin{eqnarray}
\label{eq:1}
N_{\overline{\nu}} = P_{th} (1+k(t)) \gamma,
\end{eqnarray}

where $\gamma$ is a constant parameter encompassing all non-varying terms ( free target proton number ($N_p$), detection efficiency ($\epsilon$), and standoff distance (L)). The term k(t) describes the deviation from proportionality between $N_p$ and $P_{th}$ due to the evolution of the fuel composition during the reactor cycle. It includes the fission fraction of each fuel isotope ($\alpha_i$), cross-section per fission ($\sigma_i$), and energy release per fission for each isotope ($E_i$). The explicit forms of $\gamma$ and k(t) are as follows:

\begin{eqnarray}
\label{eq:2}
\gamma = \frac{N_p \epsilon \sigma_5}{4 \pi L^2 E_5 } , \quad k(t) =  \frac{ \sum_{i=5,8,9,1} \alpha_i(t)(\frac{\sigma_i}{\sigma_5}-\frac{E_i}{E_5})}{ \sum_{i=5,8,9,1} \alpha_i(t) (\frac{E_i}{E_5})} 
\end{eqnarray}

In a reactor monitoring application, the fission rate of each isotope is estimated by measuring the antineutrino rate or energy spectrum. Since commercial reactors generally operate at constant thermal power, the change in the detected spectrum is directly related to the change in the mass fraction of each fuel isotope.

Several projects are carried out worldwide for reactor monitoring applications \cite{Battaglieri,Oguri1,Oguri2,Mulmule,Kashyap,Nucifer,Angra,PatrickHub}. Among these, highly segmented cubic-meter scale design types such as Cormorad \cite{Battaglieri}, PANDA \cite{Oguri1,Oguri2}, and ISMRAN \cite{Mulmule,Kashyap} have some advantages for safeguard application. These detectors use non-flammable plastic scintillator and thus are able to be operated safely at a desired distance from the reactor. And more importantly, they can be operated at aboveground due to its high segmented structure. 

In our previous paper, we compare existing detectors mentioned above with corresponding two new designs \cite{Kandemir1}: Hexagonal shaped packing (HSP) and Rectangular shaped packing (RSP). The main distinguishing feature of the HSP and RSP from the existing detectors is to employ hexagonal scintillator bars and light guides to form the entire detector. With this novel approach, we obtain the tightest possible arrangments of equal-sized detector units. This leads to minimizing the total surface area on which PMTs are mounted and thus reduces the number of PMTs required to read out a given detector volume.  As an example, HSP achieves the same light collection efficiency with Panda using 9$\%$ fewer PMTs (182 vs 200) in spite of having slightly larger active volume (1.02 vs 1.00 $m^{3}$). Moreover, the most important design parameter, the neutron capture time, is reduced by 8$\%$ compared to Panda\footnote{ The other superior aspects and the numerical results of the HSP are presented in detail in our previous study \cite{Kandemir1}. }. For this reason, we plan to construct HSP for monitoring Akkuyu NPP core. Current research focuses on the analysis of the antineutrino event data by reconstructing the antineutrino event in the HSP with the help of Monte Carlo based Geant4 simulation package \cite{Agostinelli}.

The rest of the paper is organized in the following orders. First, a brief description of the proposed detector is given. Second, the energy response of the detector to IBD event is investigated and the method used to analyze the IBD event data is introduced. Third, a set of event selection criteria is presented to precisely tag the antineutrino event and to reject background. Lastly, the detector response to fast neutrons are examined.


\section{Detector description }
\label{section2}
\begin{figure}[!htb]
    \centering
    
        \includegraphics[width=0.75\linewidth]{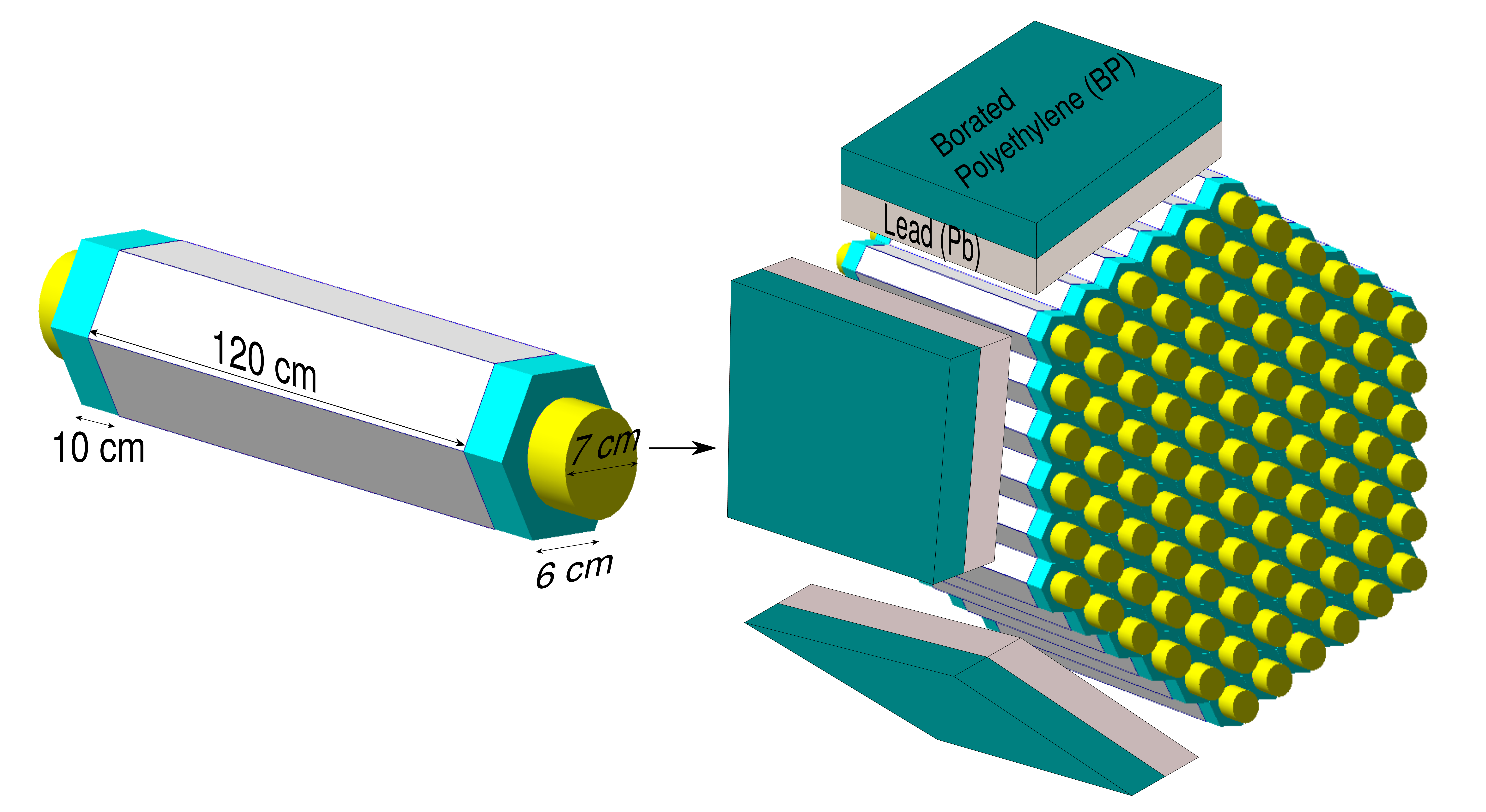}
        
    \caption{ The proposed segmented detector for monitoring nuclear reactor core activity.   }
    \label{fig:fig1}
\end{figure}

This section presents the general design of the neutrino detector planned for monitoring nuclear reactor core activity and the detection method of reactor antineutrinos. The detector is composed of identical units. Each unit has a hexagonal plastic bar (EJ-200, ELJEN Technology \cite{Eljen}) with a side length of 6 cm and a height of 120 cm, two light guides with a side length of 6 cm and a height of 10 cm, and two PMTs (9265B, ET Enterprises \cite{Et}). Plastic bars are coupled at both ends via light guides to PMTs with the help of optical cement (EJ-560). Plastic bars and light guides are wrapped with aluminized mylar film to improve the light collection. The bars are also wrapped with gadolinium-coated mylar film to enhance the neutron capture efficiency. The entire detector is formed by assembling 91 such units together into a hexagonal pattern. The detector is surrounded by passive shielding made of lead and borated polyethylene. Fig. $\ref{fig:fig1}$ depicts a schematic representation of the detector. 

The dimension of each unit is optimized considering the light collection efficiency, the neutron capture time as well the cost. The number of units is adjusted such that the active volume is around one cubic meter.

\begin{figure}[!htb]
    \centering
    
        \includegraphics[width=0.5\linewidth]{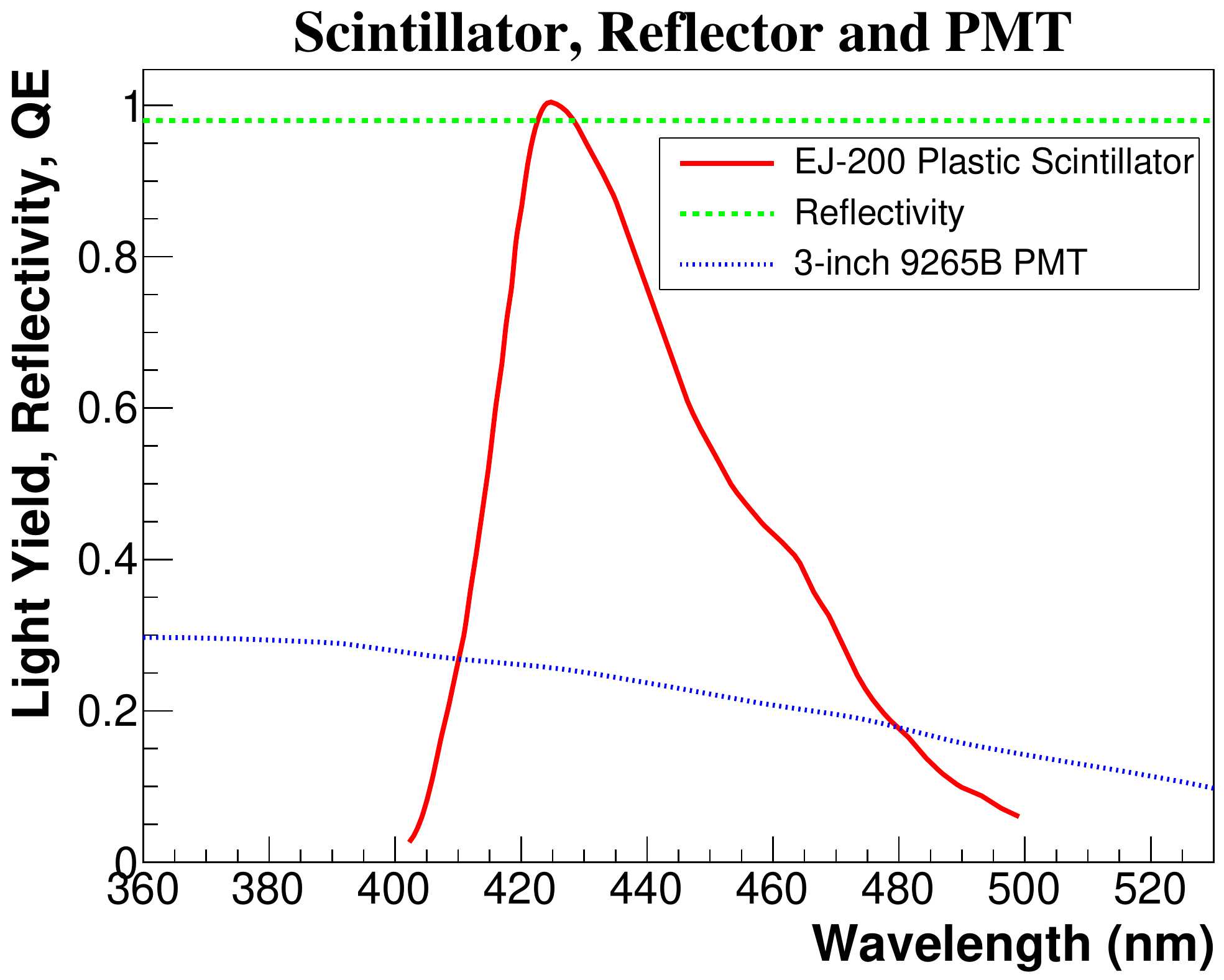}
        
    \caption{ Emission spectrum of scintillator, quantum efficiency of PMT, and reflectivity of reflector.  }
    \label{fig:fig2}
\end{figure}

The important optical parameters used in the simulation and some general features of the detector are listed in table $\ref{table:1}$. Emission spectrum of plastic scintillator, quantum efficiency of PMT, and reflectivity of the reflector are shown in Fig. $\ref{fig:fig2}$ as a function of photon wavelength. How these parameters are implemented in the simulation is explained in detail in Geant4 User's Guide \cite {App} and a study of ref. \cite{Nilson}.

\begin{table}[!htb]
    \center 
    \caption{Detector properties }
 \label{table:1}
\begin{tabular}{  p{9.2cm} p{2.1cm}  }

\hline
\quad \quad \quad  Optical properties   \\
\hline
EJ-200 Scintillator \\
Scintillation Efficiency ($\frac{photons}{MeVe^{-} }$) &  10,000  \\
Wavelength of maximum emission ($nm$) &  425  \\
Optical attenuation length ($cm$) & 380  \\
Density($\frac{g}{cm^3}$) & 1.023   \\
Refractive index & 1.58   \\
\hline
Refractive index of EJ-500 optical cement & 1.57 \\
Refractive index of light guide & 1.50 \\
Photocathode active radius (cm) &  3.5 \\
\hline
  
\quad \quad \quad Detector general properties \\
\hline
Free proton number per $cm^3$ (x$10^{22}$ ) & 5.28 \\
Detector mass (kg) & 1045\\
Detector volume ($m^3$) & 1.02 \\
Gd conc. ($\%$, w/w)  & 0.18 \\
PMT number & 182 \\

\hline
\end{tabular}
    
\end{table}

The detector uses inverse beta decay (IBD) process for $\overline{\nu}$ detection, where a $\overline{\nu}$ interacts with a proton of the scintillator target, creating a positron and a neutron : $\overline{\nu}+p \rightarrow e^+ + n $. The positron promptly deposits its energy via ionization and annihilates into two 511-keV $\gamma$-rays, producing/giving a prompt signal. The neutron thermalizes through elastic scattering off protons and then is captured by gadolinium or a hydrogen nucleus producing a delayed signal with the energy of $\approx$ 8 MeV or $\approx$ 2.2 MeV. The time-correlated detection of prompt positron signal and delayed neutron capture signal tags the $\overline{\nu}$ event and provides powerful background rejection. 
  
Due to the use of plastic scintillator bars in the proposed detector, pulse shape discrimination technique, required for distinguishing neutron and gamma-ray signals, is not effective. However, since the detector is designed as highly segmented and a neutron capture agent is employed, neutron capture events can be separated from backgrounds \cite{Mulmule}. Each passing track forms a specific hit pattern and an energy deposition profile in the detector. Analysis of hit patterns allows us to select $\overline{\nu}$ signal and reject background. For this type of detector, an additional veto counter may not be required since cosmic muons can be identified by its relatively large energy deposition in the cells of the detector along a line. 

In addition, since the proposed detector does not contain flammable materials such as a liquid scintillator, it can be operated safely on a location in close proximity to the reactor without posing any danger for reactor safety \footnote{Although the detection equipment is non-flammable, it may burn anyway at high temperatures.}. Another advantage of the proposed detector is that it can be transported and operated inside a compact vehicle.

\subsection{Energy resolution}

A detail optical photon transportation model is performed to determine the energy resolution of the proposed detector. Many parameters affecting the light collection efficiency of the detector are taken into consideration such as reflectivity of the reflector, reflector type and its applying method onto scintillator surface, and the degree of scintillator surface roughness. The impact of these parameters and how they are used in the simulation are available in our previous study \cite{Kandemir2}.

For optical photon production, 1 MeV positron is created randomly in a position inside the active volume of the detector in each event. As the positron moves, scintillation photons are produced along the track. These generated photos are then tracked throughout the detector volume until they are absorbed or detected.

\begin{figure}[!htb]
    \centering
     \includegraphics[width=0.45\linewidth]{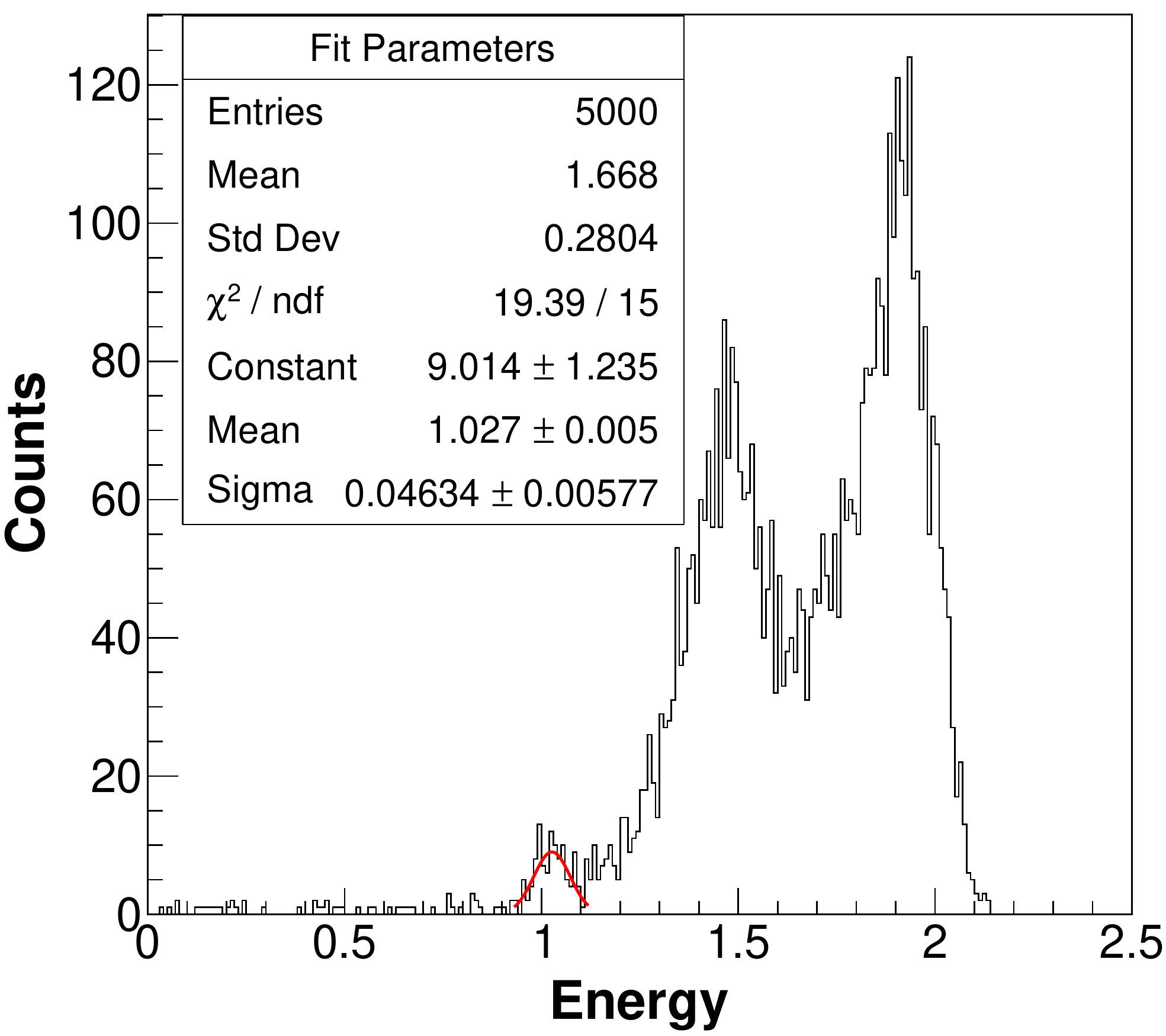}
       \caption{ 1 MeV positron energy spectrum.  }
       \label{fig:fig3}
\end{figure}

The average light collection and detection efficiency of the proposed detector are found to be $35\%$ and $8\%$, respectively. Fig. $\ref{fig:fig3}$ shows the reconstructed energy spectrum of 1 MeV positron. The full-width half-maximum (FWHM) positron energy resolution at 1 MeV is estimated to be $9\%$ (by fitting a Gaussian function) only considering the fluctuation of light generation, collection, and detection process. The final process photoelectron collection and multiplication are ignored in the simulation.

\section{Detector simulation}

\begin{figure}[!htb]
    \centering
   
        \includegraphics[width=0.46\linewidth]{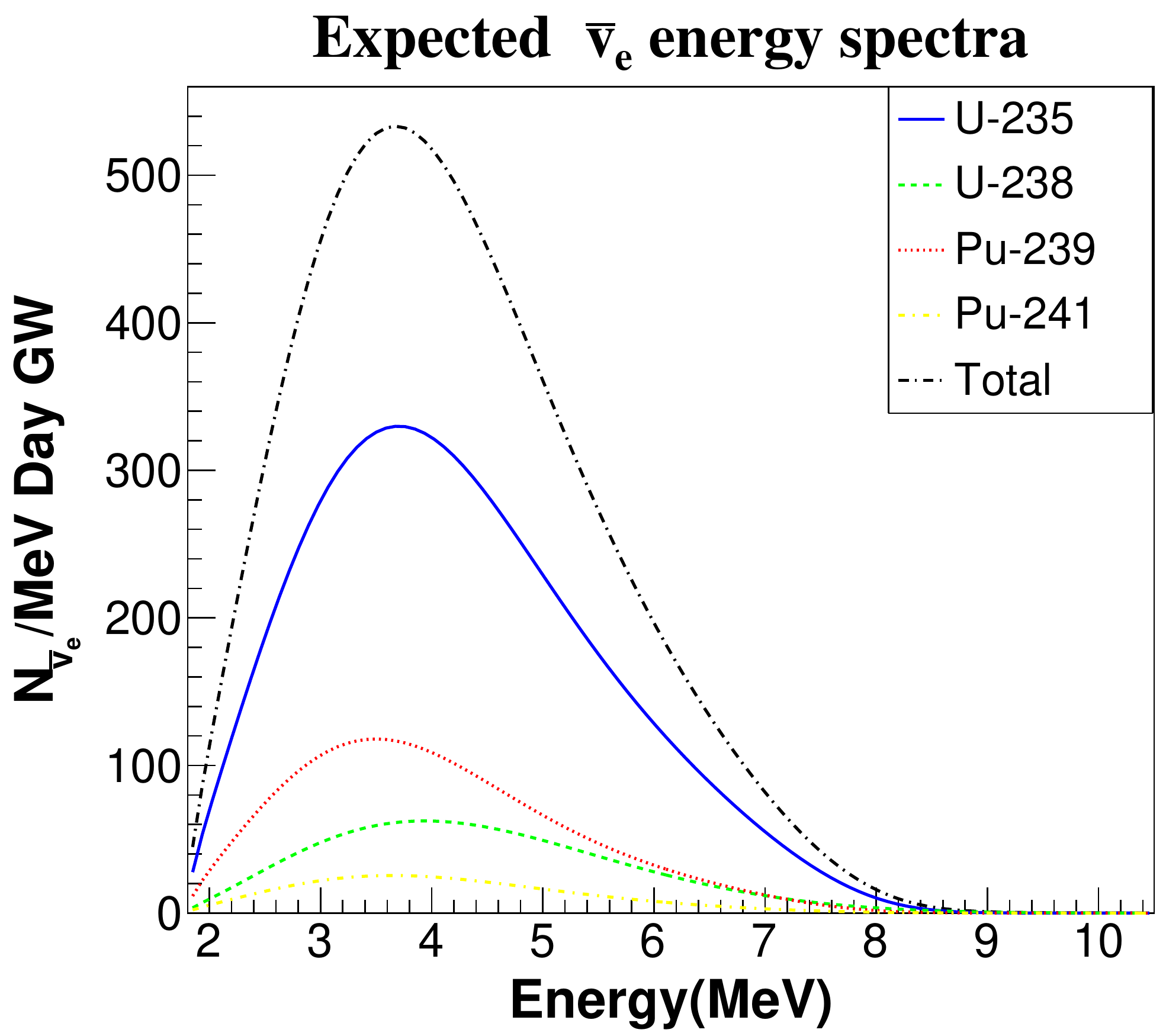}
        \caption{Expected $\overline{\nu}$ interaction rate for each fuel isotope. The distance between the detector and the reactor core is taken as 20 m in the calculation.  }
        \label{fig:fig4}
\end{figure}

Monte Carlo based Geant4 simulation toolkit (version 4.10.4) is used to simulate the IBD event in the proposed detector. Antineutrino interaction vertex is created in a random point inside the active volume of the detector with the $\overline{\nu}$ energy selected from the expected antineutrino spectrum (total spectrum in Fig. $\ref{fig:fig4}$). The initial energy of the IBD products positron and neutron is derived from the IBD kinematics \cite {Vogel}. The simulation starts with the generation of positron and neutron and ends with the detection of ensuing scintillation photons. All the particles produced during the simulation are tracked throughout the detector volume to estimate the deposited energy in each cell of the detector considering the time of energy deposition.

The QGSP-BERT-HP physics package \cite{Berthp}, which includes both the advanced hadronic (high precision neutron transport model ) and standard electromagnetic physics process, is utilized for the primary and secondary particle interactions. For scintillation photon,  Geant4 offers two optical simulation models: the UNIFIED \cite {Unified} and the GLISUR model. The UNIFIED model is chosen for detailed scintillator surface wrapping.  A total of 50.000 IBD event is simulated for each run to achieve sufficient statistics. The simulated data are then analyzed by the ROOT framework \cite {Root}.

\subsection{Method}

When searching for antineutrinos with the delayed coincidence technique, two types of background events are encountered: the correlated and the uncorrelated. The correlated background consists of a single physical process that results in two time-correlated signals, as in the antineutrino event. Fast neutrons produced from cosmic muons are the main source of this type of background. The second kind happens when an independent positron-like and neutron-like event randomly occurs within the time window of the delayed coincidence search. The true IBD signal can be discriminated from the background by establishing proper event selection criteria.

Antineutrino event selection criteria are based on the following variables defined for both prompt and delayed signal of the IBD interaction: total energy deposited in all cells of the detector ($E_{total}$), the number of triggered cells ($N_{hit}$), and the energy of the four cells with the highest energy deposition ($E_{1st}$, $E_{2nd}$, $E_{3th}$ and $E_{4th}$). The width of the delayed coincidence time window, which is adjusted according to the time interval between the prompt and delayed signal ($\Delta$T), is also an important selection criterion. The distributions of these variables are obtained from the IBD event data. By examining the distribution of these variables, a set of selection criteria is developed to precisely tag the $\overline{\nu}$ event and to discriminate it from the background. 

The selection efficiency for each selection criteria is calculated from simulations and defined as the ratio of the number of events satisfying the selection conditions to the total number of events. For the uncertainty in each selection efficiency, it is assumed that systematic uncertainties are caused only by uncertainties in the simulation model. The relative uncertainty in the simulation model is estimated to be less than 20$\%$ in ref. \cite{Oguri2}\footnote{In this study, in order to estimate the uncertainties in the simulation model, dedicated experiments with different radioactive sources are performed and the detection rates are compared between the simulation and experiment.}. Statistical uncertainty is negligible since a sufficient number of events are generated.

To achieve spectral uniformity among different bars, a threshold energy of 0.2 MeV is applied to the energy deposited in each bar of the detector as a pre-selection criterion.

\subsubsection{Prompt signal}

In a typical prompt event, positron usually deposits its energy within the same cell which it is created in. The two 511 keV gammas, which are created from the positron annihilation, mostly escape from the cell and leave their energy in the neighboring cells. Since the energy deposition of these two processes occurs promptly, the sum of the two energies is seen as a single event in the detector and defines the prompt signal.  

\begin{figure}[!htb]
    \centering
    \subfigure[  ]
    {
        \includegraphics[width=0.46\linewidth]{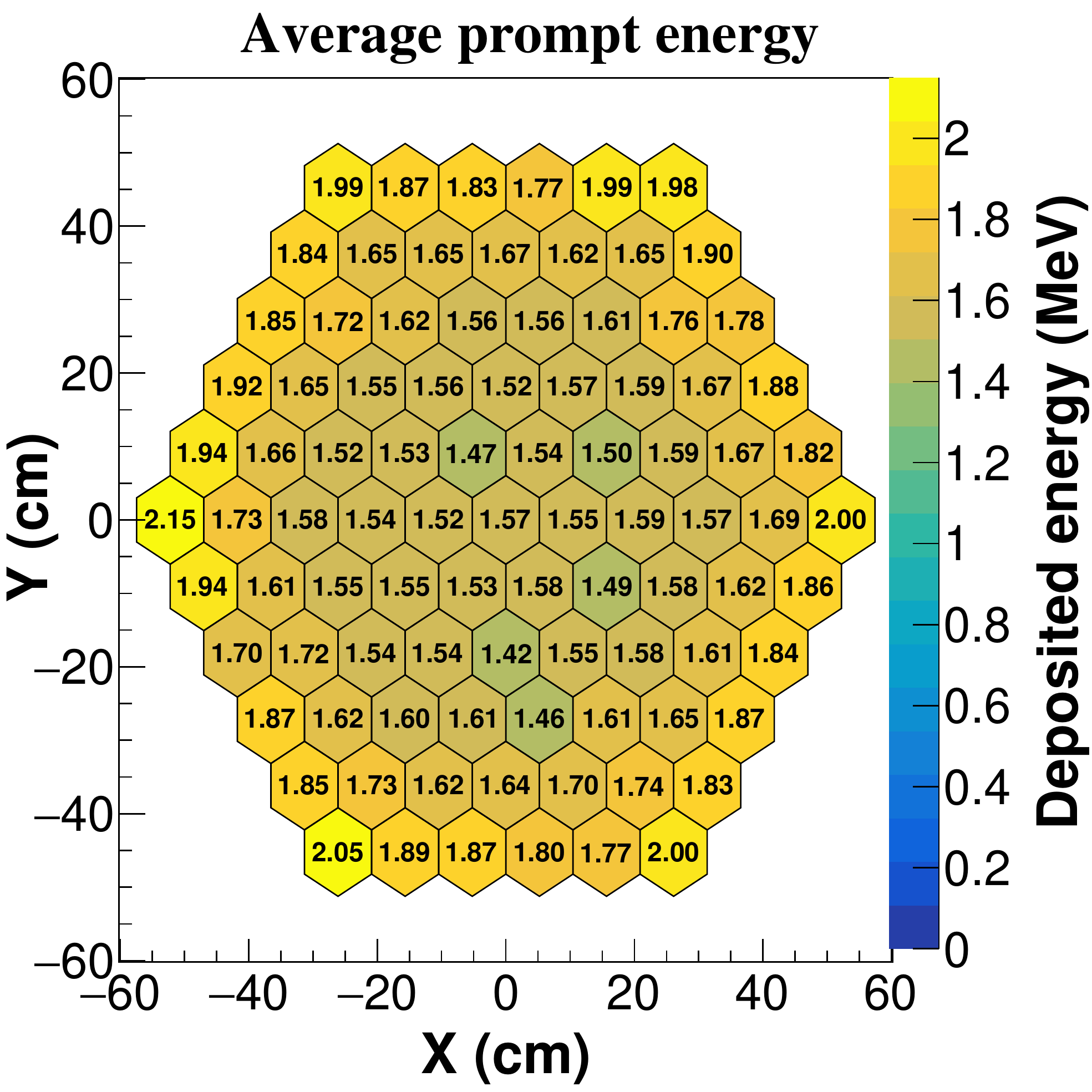}
        \label{fig:fig5a}
    }
    \quad
   \subfigure[ ]
    {
       \includegraphics[width=0.46\linewidth]{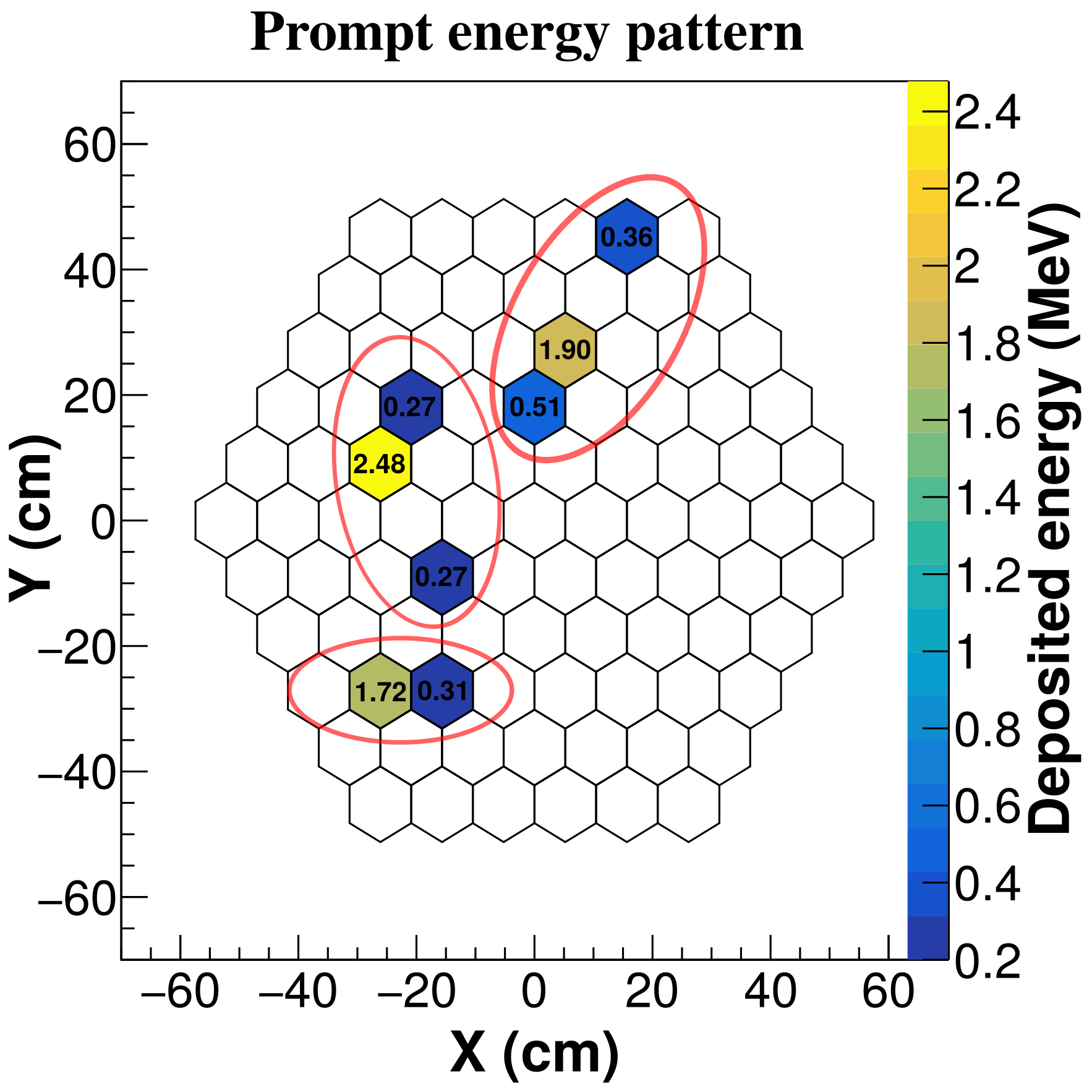}
        \label{fig:fig5b}
    }
    \caption{ (a) The mean energy deposited in each cell of the detector for prompt events. (b) Energy deposition profile of three different prompt events. Each red circle represents an independent event. }
    \label{fig:fig5}
\end{figure}

To estimate how much energy is deposited in each cell of the detector in prompt events, we calculate the mean energy deposited in each cell by averaging over all events in which there is an accumulation of energy in that cell. Fig. $\ref{fig:fig5a}$ shows the mean energy deposited in each cell of the detector in prompt events. The average energy deposition in the cells of the outer part of the detector is slightly higher than the cells in the interior. The reason for this is that the gammas produced from positron annihilation in the outer cells are more likely to escape from the detector with respect to the inner cells. 

\begin{figure}[!htb]
    \centering
    
        \includegraphics[width=0.46\linewidth]{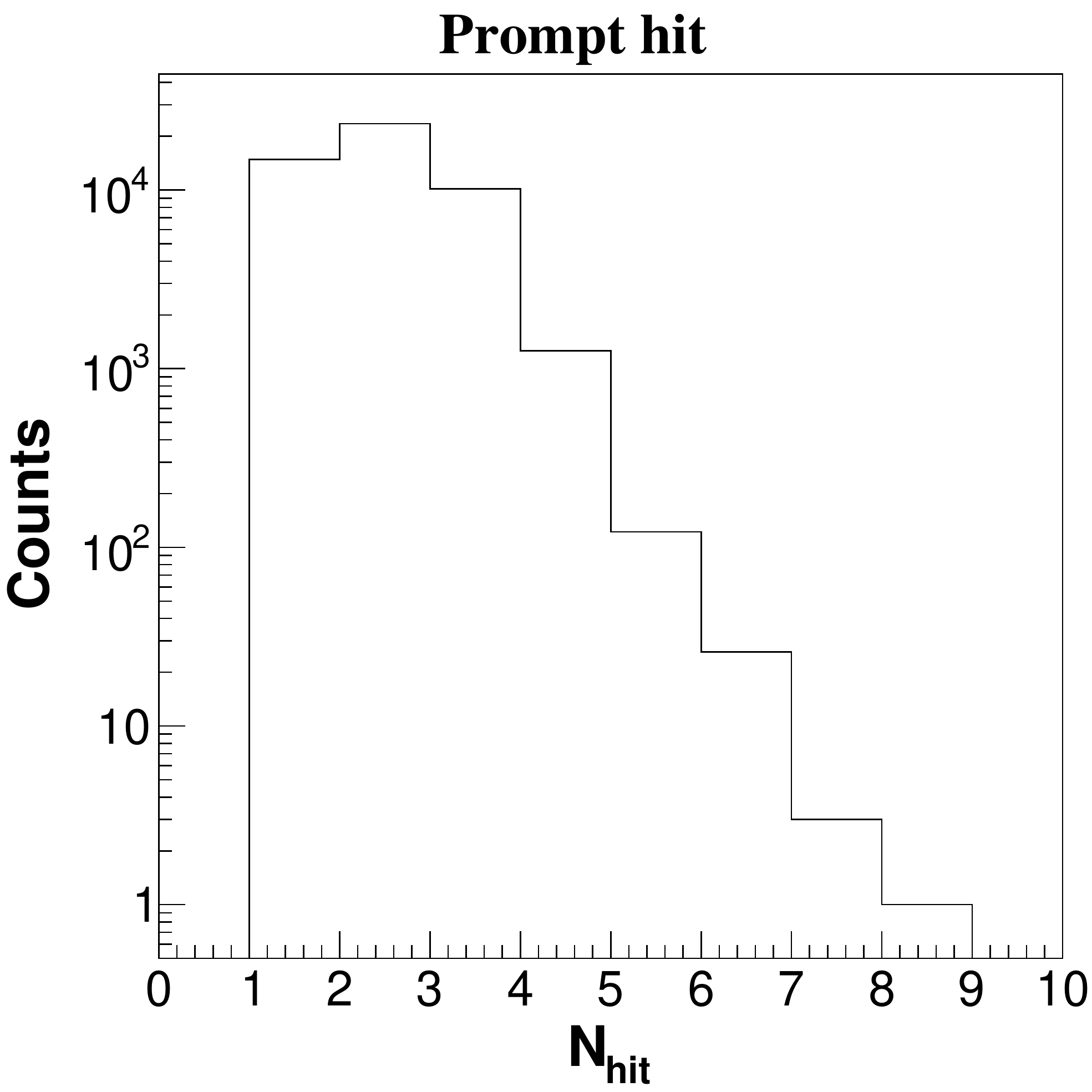}
        \caption{The triggered number of detector cells in prompt events.}
        \label{fig:fig6}
\end{figure}

To reveal more characteristics of the prompt event, we look at the individual events. Fig. $\ref{fig:fig5b}$ shows energy deposition pattern of the prompt signal in the proposed detector for three different events. As in all three cases in Fig. $\ref{fig:fig5b}$, the prompt event is characterized as a condition in which a significant amount of energy is left in a cell and a relatively low energy accumulation to the cells around it. The hit multiplicity of the prompt event comes from the two annihilation gammas. Indeed, positron deposits its energy in a single cell in 83$\%$ of the events and in two cells in 17$\%$ of the events. Therefore, all cases of $N_{hit}>$2 certainly includes the energy deposition of annihilation gammas. Fig. $\ref{fig:fig6}$ shows the number of triggered detector cells in prompt events. It is inferred from Fig. $\ref{fig:fig6}$ that a single cell is triggered in 29$\%$ of the prompt events, two cells in 45$\%$ and three cells in 20$\%$.

\begin{figure}[!htb]
    \centering
    \subfigure[  ]
    {
        \includegraphics[width=0.46\linewidth]{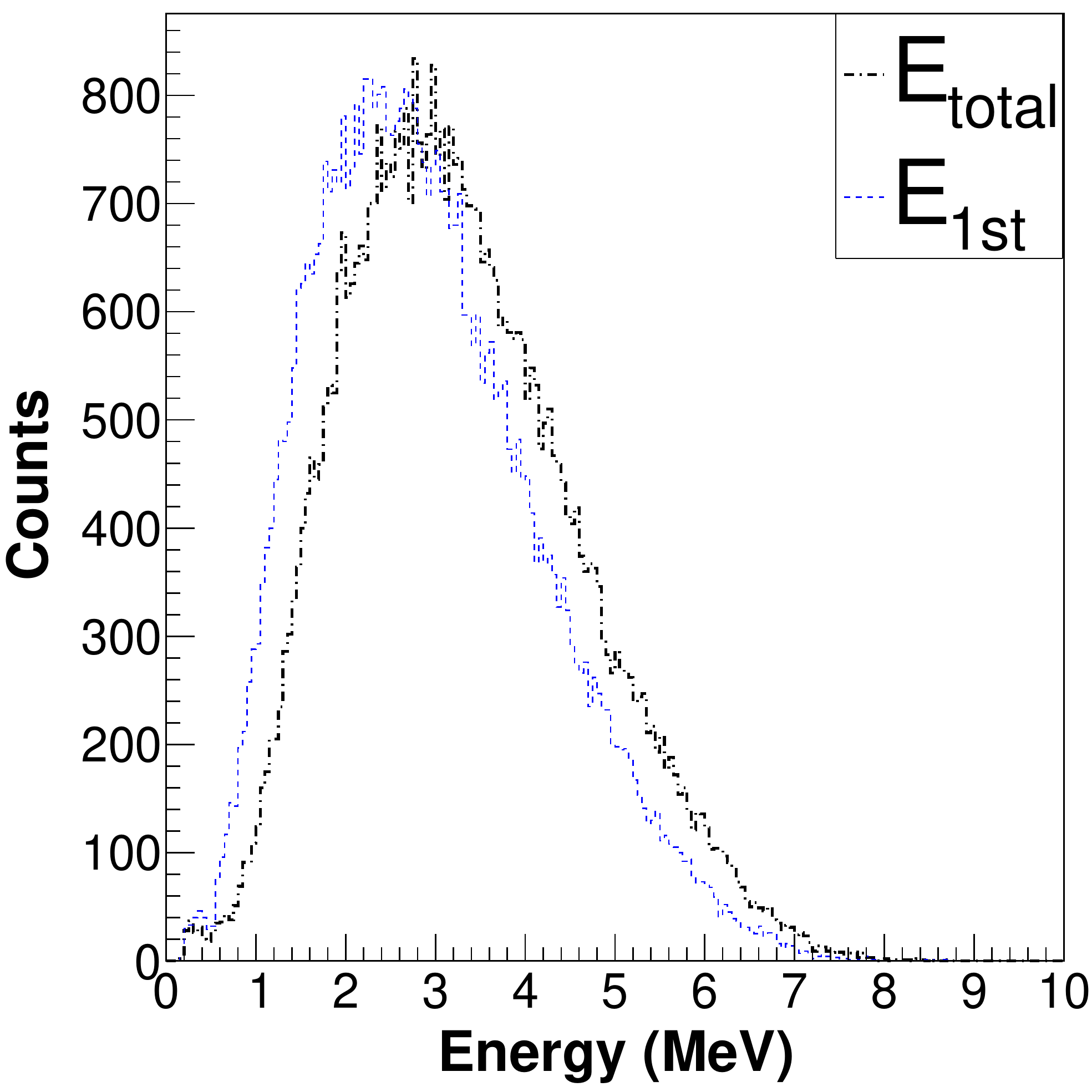}
        \label{fig:fig7a}
    }
    \quad
    \subfigure[  ]
    {
       \includegraphics[width=0.46\linewidth]{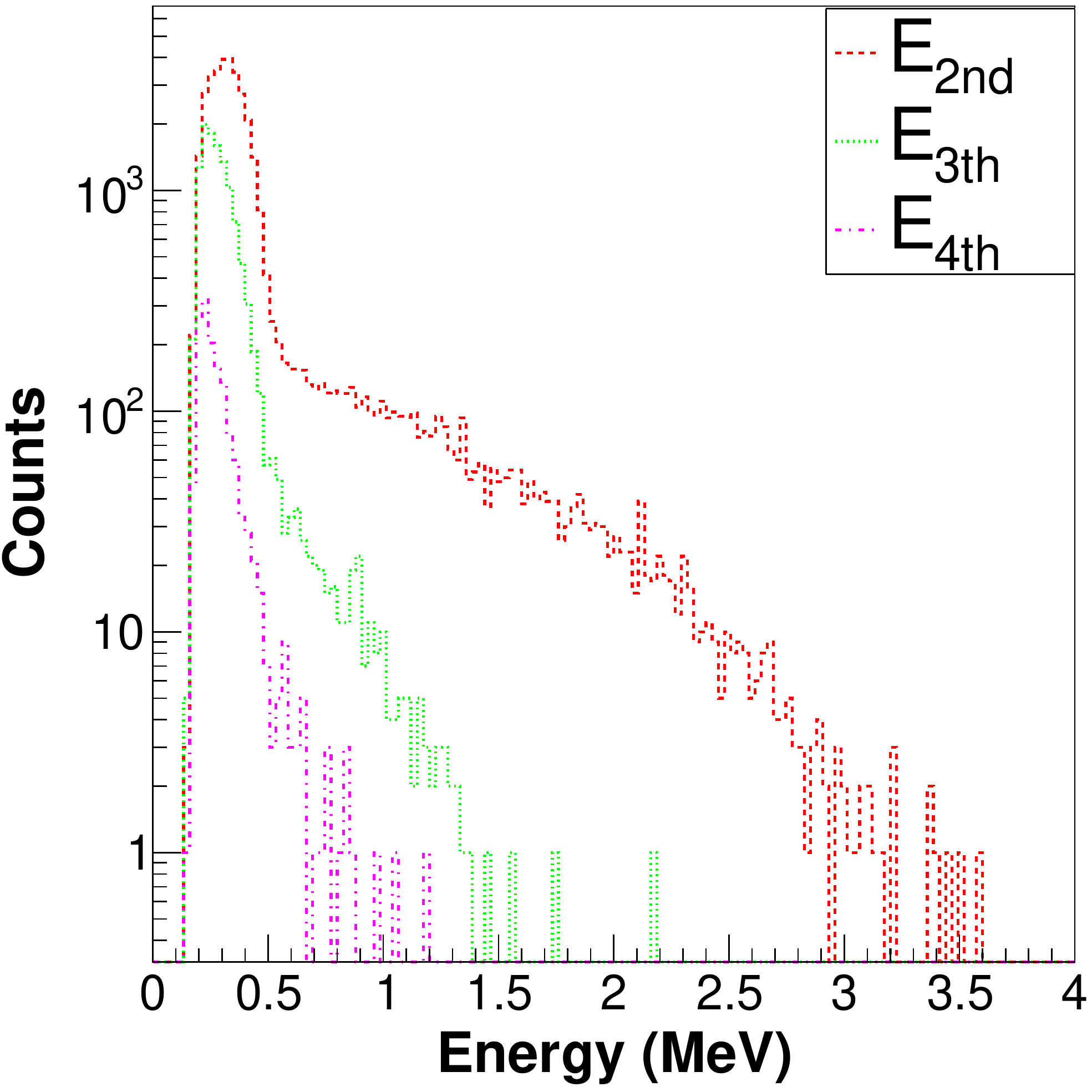}
        \label{fig:fig7b}
    }
    \caption{ Total deposited energy of all cell of the detector ($E_{total}$) and the first four cells where the highest energy deposited ($E_{1st}$, $E_{2nd}$, $E_{3th}$ and $E_{4th}$).  }
    \label{fig:fig7}
\end{figure}

The energy distribution of the prompt energy ($E_{total}$) and the cell with the highest energy deposition ($E_{1st}$) are shown in Fig. $\ref{fig:fig7a}$. We set the threshold energy $E_{totat}$ to 2.5 MeV to eliminate background gammas arises from the ambient and reactor related gamma-ray. Also we set the upper limit of the energy deposited in any cell to 6 MeV.

Fig. $\ref{fig:fig7b}$ shows energy distributions of the cells where the second, third, and fourth highest energy is deposited. The peak at 324 keV (black lines) comes from the Compton edge of 511 keV annihilation gammas. Including the events where one of the gammas deposits all of its energy, we choose an upper threshold of 520 keV for the second condition to tag the annihilation gammas.

\subsubsection{Delayed signal}

The neutron emerged from the IBD reaction thermalizes in the active volume of the detector by making elastic collisions with the scintillator nuclei. The thermalized neutron is then captured by mostly a gadolinium or rarely a hydrogen nucleus. The excited nucleus then emits gamma rays. Energy deposition of these gamma-rays in the detector produces the delayed signal.

The energy and number of gammas released from the neutron capture depend on the nucleus type that captures neutrons. It is observed that 67$\%$ of the IBD neutrons are captured on gadolinium isotopes and 18$\%$ are captured on hydrogen. A small fraction of neutrons is also captured on Carbon (0.3$\%$). In the case of capture by gadolinium, a cascade of gamma rays is emitted with different total energy depending on the isotope type.  On the contrary, a single gamma of 2.2 MeV energy is released in the case of capture by hydrogen. The results are presented in detail in Table $\ref{table:2}$.

\begin{table}[!htb]
\center
\caption{Neutron capture comparison. $<N_{\gamma}>$ is the mean number of gammas emitted by the nucleus that captures the neutron, and $<E_{\gamma}>$ is the average total energy of emitted gammas.}
\begin{tabular}{llll}
 Nucleus & Fraction($\%$) & $<N_{\gamma}>$   & $<E_{\gamma}>$  \\
 \hline
 Gd-155 & 13  & 11.1 & 8.4 \\
 Gd-157 & 56 & 2.5  &  7.9 \\
 Gd-156 & 0.02 & 4.0 & 6.4 \\
 Gd-158 & 0.02 & 4.6 & 5.9 \\
 \hline
 Hydrogen & 20  & 1.0  & 2.2 \\
 \hline
 Carbon & 0.3 & 1.4 &  4.9 \\
 \hline
 Escape & 10 & - & - \\
 \hline
\end{tabular}
\label{table:2}
\end{table}

\begin{figure}[!htb]
    \centering
    \subfigure[  ]
    {
        \includegraphics[width=0.46\linewidth]{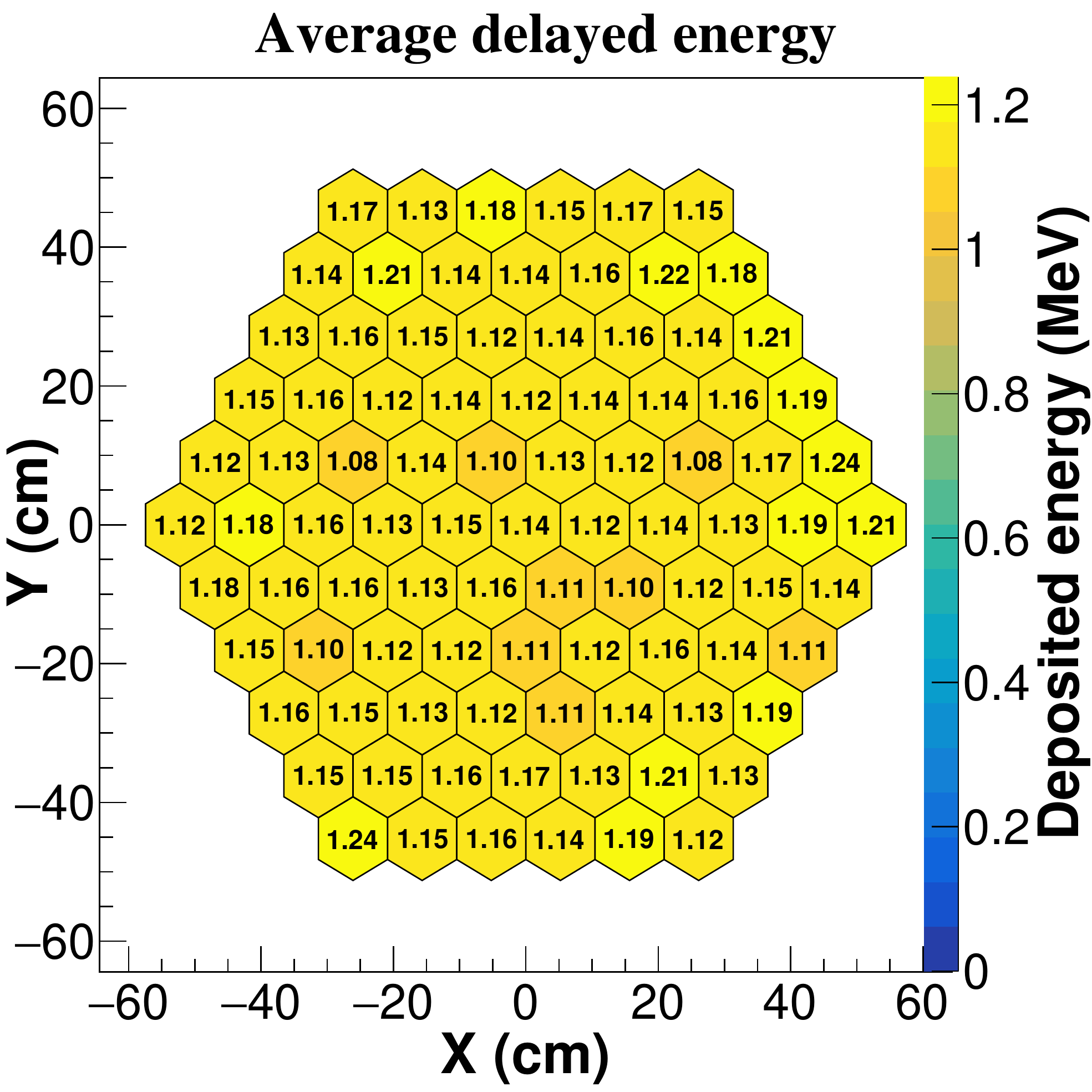}
        \label{fig:fig8a}
    }
    \quad
    \subfigure[ ]
    {
       \includegraphics[width=0.46\linewidth]{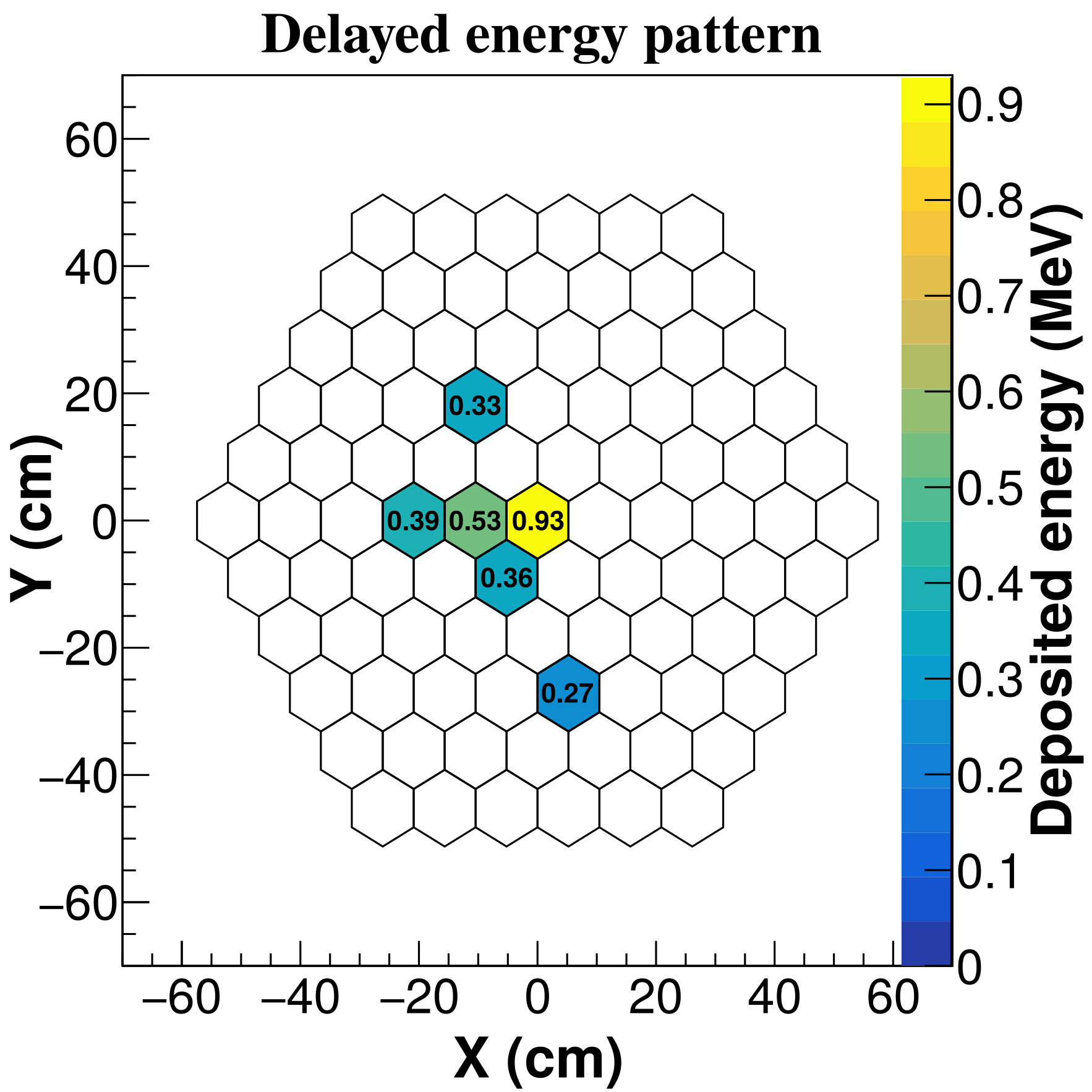}
        \label{fig:fig8b}
    }
    \caption{ (a) The mean energy deposited in each cell of the detector for delayed events. (b) An example of energy deposition pattern of a delayed event.   }
    \label{fig:fig8}
\end{figure}

\begin{figure}[!htb]
    \centering
    \subfigure[  ]
    {
        \includegraphics[width=0.46\linewidth]{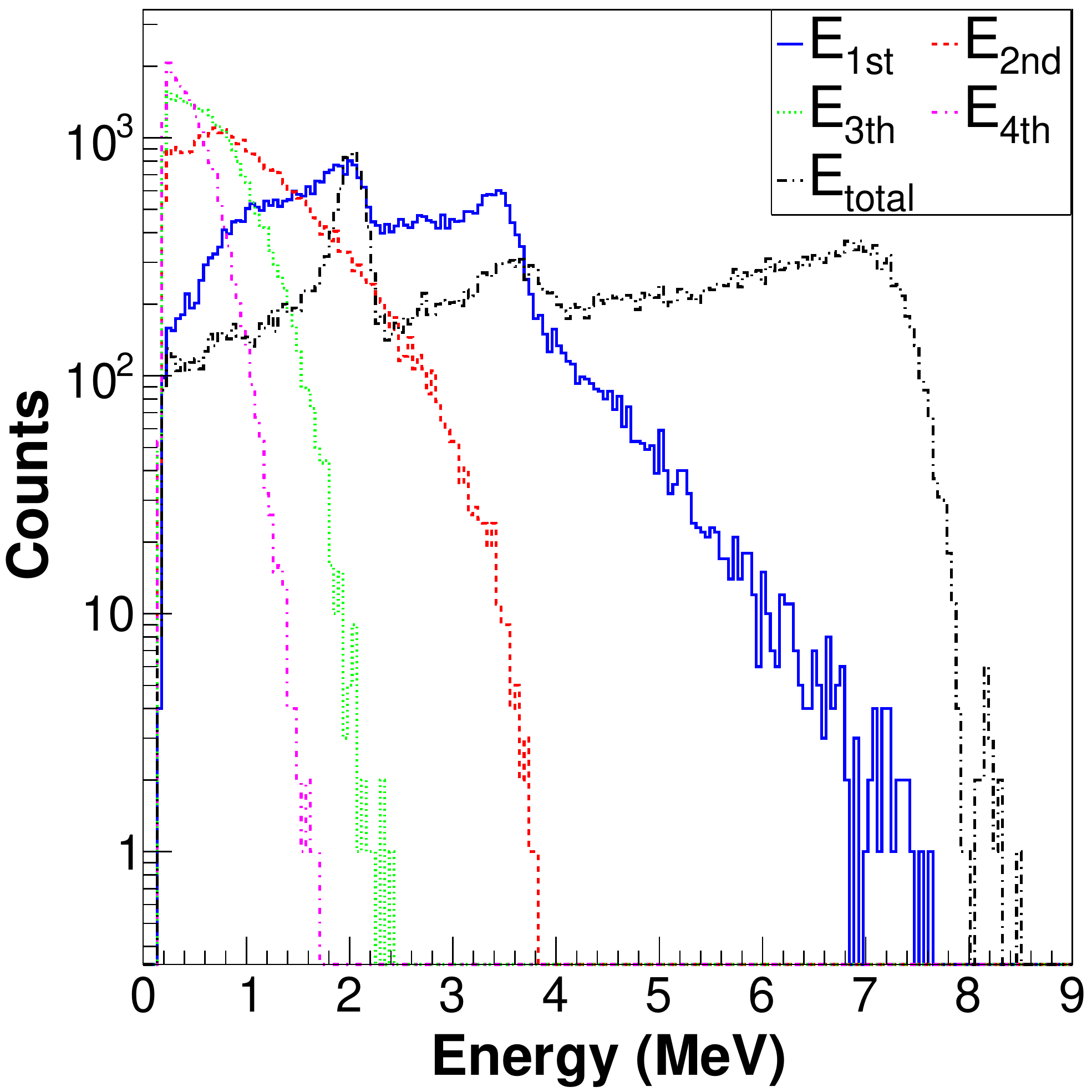}
        \label{fig:fig9a}
    }
    \quad
    \subfigure[ ]
    {
       \includegraphics[width=0.46\linewidth]{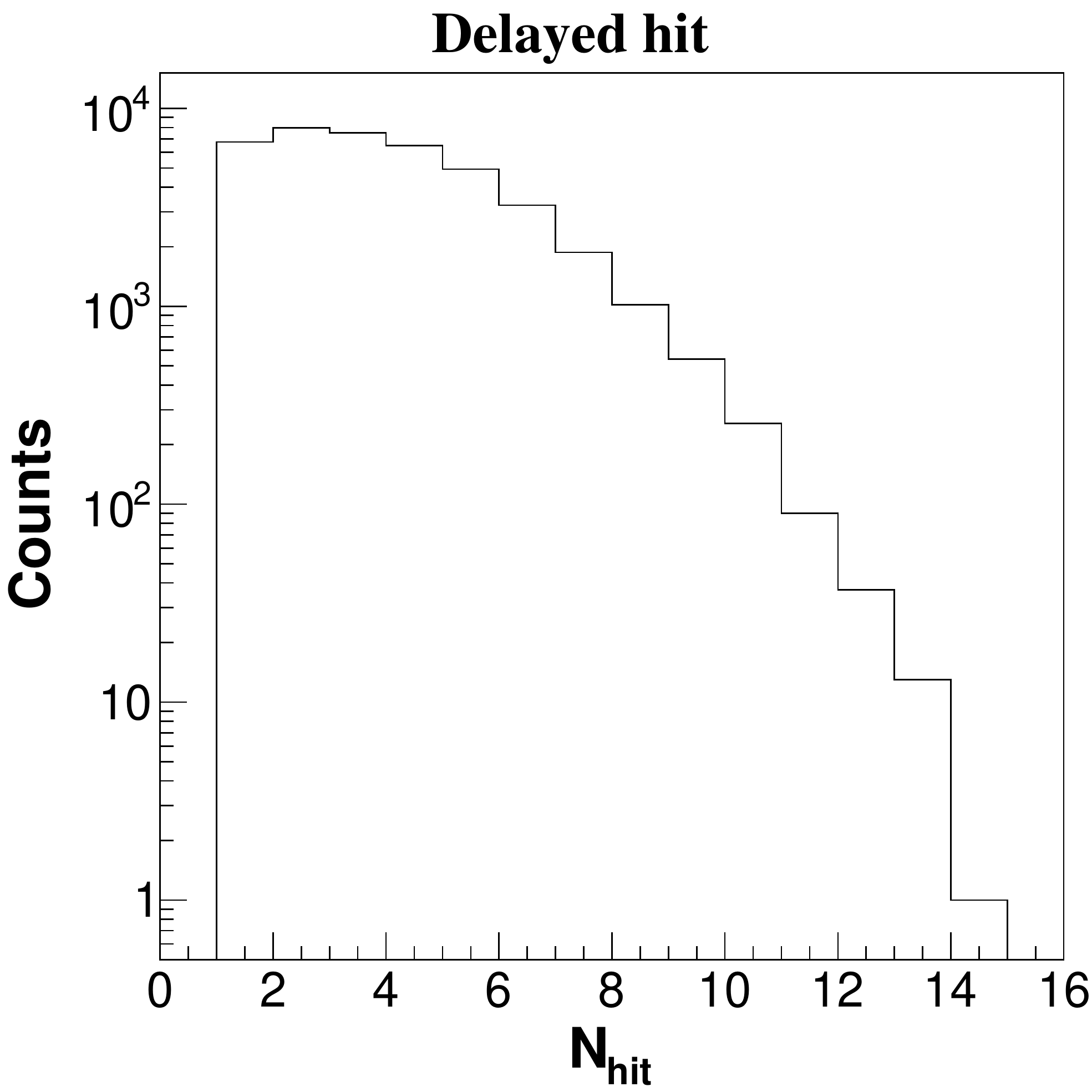}
        \label{fig:fig9b}
    }
    \caption{ (a) The energy distribution of the four highest energy deposition cells for delayed events. (b) The triggered number of detector cells for delayed events.  }
    \label{fig:fig9}
\end{figure}

Fig. $\ref{fig:fig8a}$ shows the average energy deposited in each cell of the detector for all delayed events, while Fig. $\ref{fig:fig8b}$ shows the energy deposition pattern of a single delayed event in the detector. As shown in Fig. $\ref{fig:fig8b}$, the delayed energy is not concentrated in a single cell as in the prompt event. On the contrary, a few cells received a considerable amount of energy. Fig. $\ref{fig:fig9a}$ shows the energy distribution of the four cells where the highest energy is deposited. Referring to Fig. $\ref{fig:fig9a}$, we set the upper limits of $E_{1st}$, $E_{2nd}$, $E_{3th}$, and $E_{4th}$ to 6, 3, 2, and 1 MeV, respectively. Fig. $\ref{fig:fig9b}$ shows how many numbers of cells are triggered in delayed events.

In addition to prompt and delayed signal efficiency, $\overline{\nu}$ detection efficiency is also affected by the time difference between the prompt and delayed energy depositions ($\Delta$T). This time interval changes according to the neutron capture time. The neutron capture time also depends on the geometry of the detector, gadolinium concentration, and how the gadolinium is distributed over the active volume of the detector. The time of capture follows an exponential distribution and the mean of the distribution is 56 $\mu$s for the proposed detector. We calculate $\overline{\nu}$ detection efficiency with the given prompt-delayed time intervals when all the selection cuts are applied. The results are reported in Table $\ref{table:3}$. 
It is estimated that $\overline{\nu}$ detection efficiency reaches to saturation nearly within 200 $\mu$s, and 10$\%$ of the $\overline{\nu}$ events can be detected within the time interval of 4-200 $\mu$s. Reaching the efficiency saturation in short time is crucial since keeping the prompt-delayed time window short reduces uncorrelated background event rates that occur within this time interval and hence increases efficiency. 

\begin{table}[!htb]
\caption{Estimated selection efficiency of each stage of antineutrino detection and background rejection. For correlated background, only rejection of fast neutrons passing 2.5 MeV $<$ $E_{total}$ $\leq$ 8 MeV (prompt), 3 MeV $<$ $E_{total}$ $\leq$ 8 MeV (delayed), and 4 $<$ $\Delta$T $<$ 200 cuts are quoted. Systematic uncertainties in the efficiencies are considered to be less than 20$\%$ based on ref. \cite{Oguri2}.  }
\center
\begin{tabular}{p{5.2cm} p{2.5cm}  p{2.0cm}  p{2.0cm} p{2.0cm} p{2.0cm} }
\hline
   & \begin{tabular}[c]{@{}l@{}}Antineutrino \\detection \end{tabular} & \multicolumn{2}{l}{\begin{tabular}[c]{@{}l@{}}Single fast neutrons\\rejection \end{tabular}}  & \multicolumn{2}{l}{\begin{tabular}[c]{@{}l@{}}Multi-neutrons\\ rejection \end{tabular}} \\
   \hline
&  & \multicolumn{3}{c}{Initial neutron spectrum} \\
\hline
 
 &     & \begin{tabular}[c]{@{}l@{}}Random\\(1-50 MeV) \end{tabular}  & \begin{tabular}[c]{@{}l@{}}Gordon \cite{Gordon} \end{tabular} & \begin{tabular}[c]{@{}l@{}}Random\\(1 keV- \\50 MeV) \end{tabular} & \begin{tabular}[c]{@{}l@{}}Gordon \end{tabular} \\
\hline
 Selection criteria & \multicolumn{5}{c}{ Efficiency ($\%$)} \\
 \hline
 \multicolumn{2}{l}{ Prompt signal} \\
 2.5 MeV $<$ $E_{total}$ $\leq$ 8 MeV & 70 & -  & -  & -  & -  \\
 1 MeV $<$ $E_{1st}$ $\leq$ 6 MeV     & 95 & 13 & 6  & 3  & 9  \\
 $E_{2nd}$ $\leq$ 520 keV             & 59 & 90 & 88 & 93 & 91 \\
 1 $<$ $N_{hit}$ $\leq$ 3             & 67 & 40 & 41 & 69 & 57 \\
 Total                                & 40 & 92 & 90 & 96 & 95 \\
 \hline
 \multicolumn{2}{l}{ Delayed signal} \\
 3 MeV $<$ $E_{total}$ $\leq$ 8 MeV   & 52 & -  & -  & -  & - \\
 0.5 MeV $<$ $E_{1st}$ $\leq$ 6 MeV   & 79 & 1  & 9  & 1  & 9 \\
 $E_{2nd}$ $\leq$ 3 MeV               & 67 & 4  & 10 & 4  & 10 \\
 $E_{3th}$ $\leq$ 2 MeV               & 51 & 14 & 27 & 12 & 27 \\
 $E_{4th}$ $\leq$ 1 MeV               & 35 & 36 & 53 & 33 & 52 \\
 3 $<$ $N_{hit}$ $\leq$ 6             & 29 & 48 & 55 & 47 & 56\\
 Total                                & 25 & 51 & 59 & 49 &  58 \\
 \hline
 \multicolumn{2}{l}{ \begin{tabular}[c]{@{}l@{}}Prompt-delayed time\\ interval ($\mu$s) \end{tabular} } \\
 4 $<$ $\Delta$T $<$ 200              & 82 & -  & -  & -  & - \\
 \hline
 Total (includes all cuts)            & 10 & 96 & 96 & 98 & 98\\
 \hline
  \multicolumn{6}{c}{ Variation of $\overline{\nu}$ detection efficiency as a function of prompt-delayed time interval } \\
  \hline
 Time window ($\mu$s) &\multicolumn{5}{l}{Efficiency ($\%$)  } \\
  \hline
  
4 $<$ $\Delta$T $<$ 50  & 6 \\

50 $<$ $\Delta$T $<$ 100  & 2.4 \\

100 $<$ $\Delta$T $<$ 150  & 1.1 \\

150 $<$ $\Delta$T $<$ 200  & 0.5 \\
200 $<$ $\Delta$T $<$ 500  & 0.3 \\
\hline 
\end{tabular}
\label{table:3}
\end{table}

For the proposed detector for Akkuyu NPP, we compute the expected number of detected antineutrinos by using equation $\ref{eq:1}$. The detector is assumed to be located in 20 m distance from the reactor. The fission fraction evolution and the energy release per fission of each isotope are taken from ref. \cite{Bemporad} and \cite{Erpf}, respectively for the calculation. Fig. $\ref{fig:fig10}$ shows the predicted evolution of detected antineutrinos over the course of the fuel cycle.

\begin{figure}[!htb]
    \centering
    
        \includegraphics[width=0.46\linewidth]{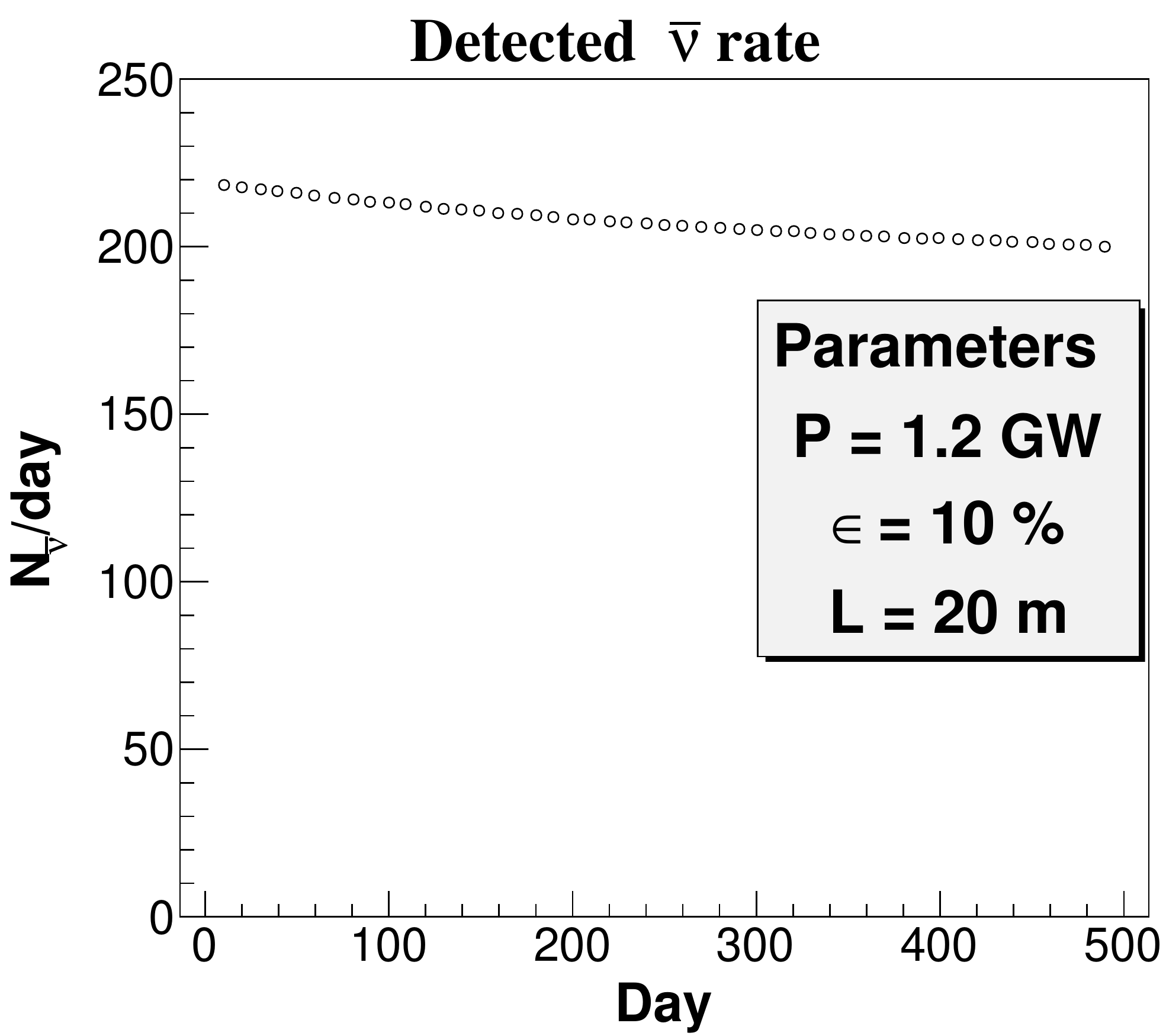}
        \label{fig:fig10}
        \caption{ The evolution of the detected antineutrino number over the course of the fuel cycle.  }  
\label{fig:fig10}
\end{figure}

\subsection{Fast neutron}

Two types of correlated background events are considered in the simulation: single fast neutrons and multi-neutrons. When a fast neutron enters the detector, it loses its energy by making elastic collisions with the protons of the scintillator. The thermalized neutron is then captured by a neutron-absorbing nucleus. The recoiled protons mimic the prompt signal while the gammas resulting from neutron capture mimic the delayed signal. On the other hand, when generated in the same cosmic shower, two thermalized neutrons can get captured by gadolinium nuclei inside the active volume of the detector. The earlier capture cannot be discriminated from the positron signal and thus appears as an antineutrino signal. Elimination of these backgrounds is the key issue for above-ground detection.

\begin{figure}[!htb]
    \centering
    
        \includegraphics[width=0.48\linewidth]{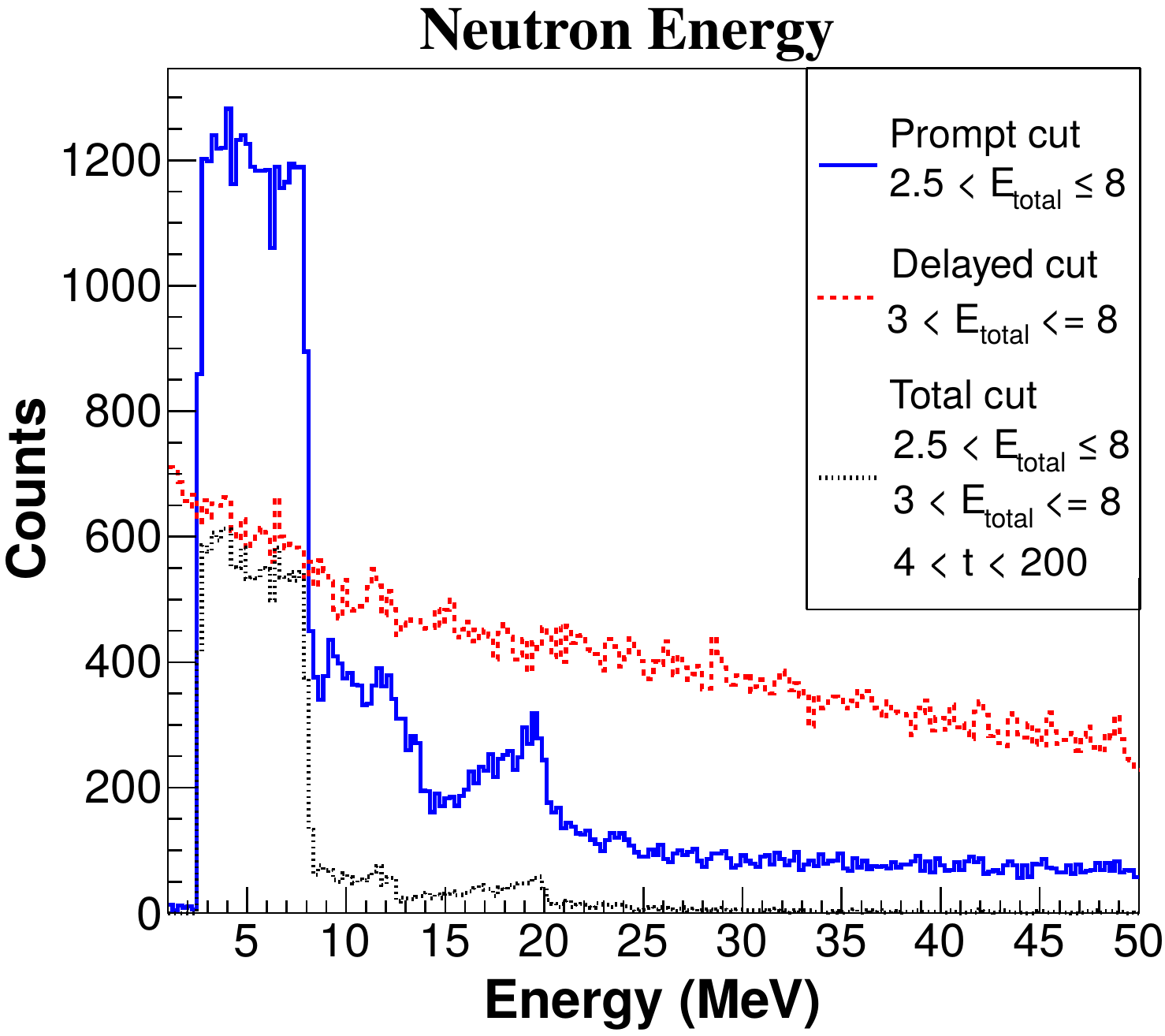}
        \caption{ The impact of initial neutron energy used as input in the simulation on the number of events that satisfy $E_{Total}$ (both prompt and delayed) and time ($\Delta$T) cuts. }  
\label{fig:fig11}
\end{figure}

To investigate which neutron energies are most likely to cause single fast neutron backgrounds, neutrons are simulated in the range of 1-50 MeV. For each event, a neutron is created at a point inside the detector with the energy chosen randomly from this energy interval and fired in any direction. The deposited energy arising from the recoiled protons (fake-prompt) and the neutron (fake-delayed) is recorded separately. Fig. $\ref{fig:fig11}$ shows the number of events passing the $E_{total}$ cuts (for both prompt and delayed signal shown in Table $\ref{table:3}$) as a function of initial neutron energy. Only 5$\%$ of the generated events pass the total cuts (black line), and 81$\%$ of these passed events come from the neutron energy range of 2.5-8 MeV. From Fig. $\ref{fig:fig11}$, it is concluded that prompt cuts are very effective in eliminating neutrons below 2.5 MeV and above 8 MeV.

\begin{figure}[!htb]
    \centering
    \subfigure[  ]
    {
        \includegraphics[width=0.46\linewidth]{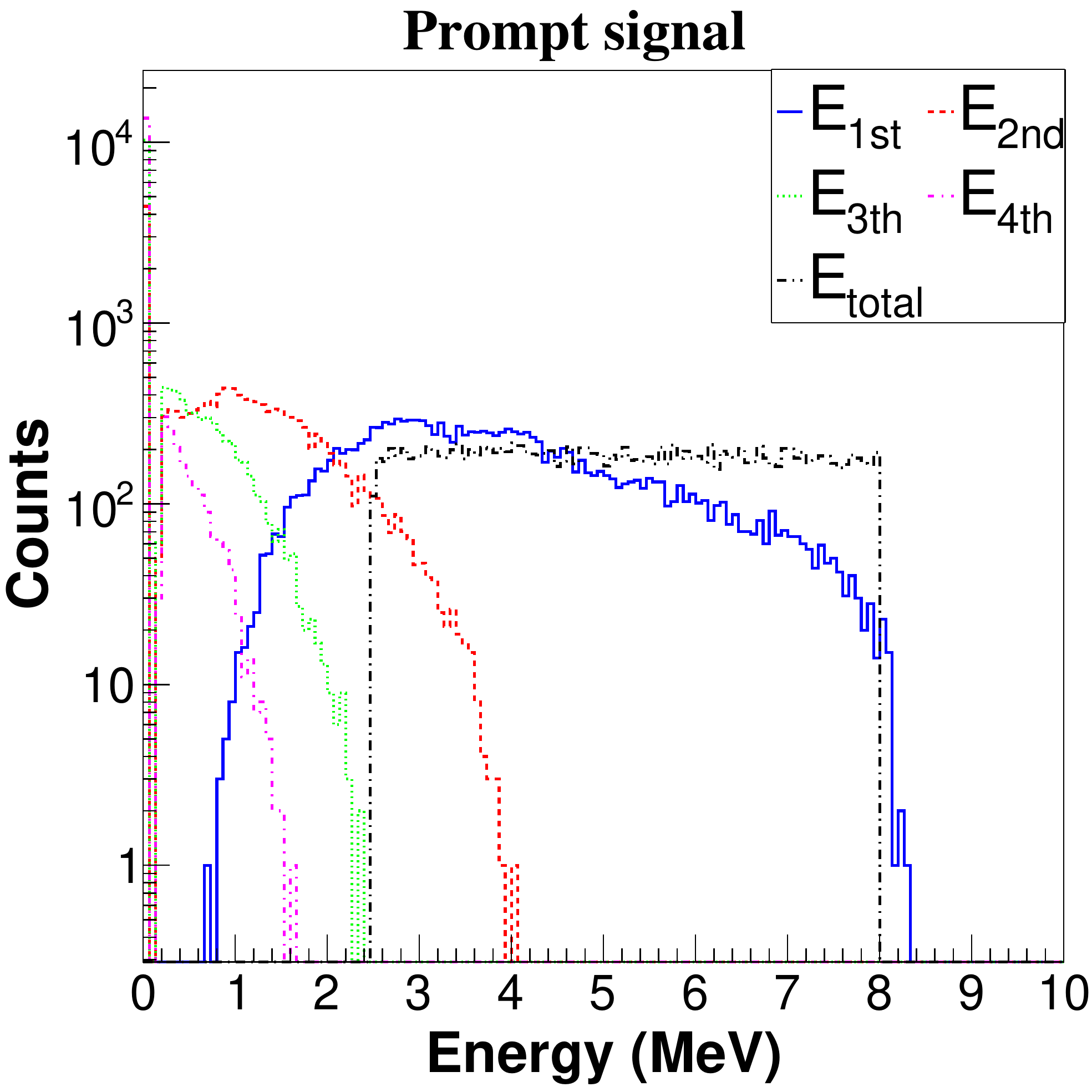}
        \label{fig:fig12a}
    }
    \quad
    \subfigure[ ]
    {
       \includegraphics[width=0.46\linewidth]{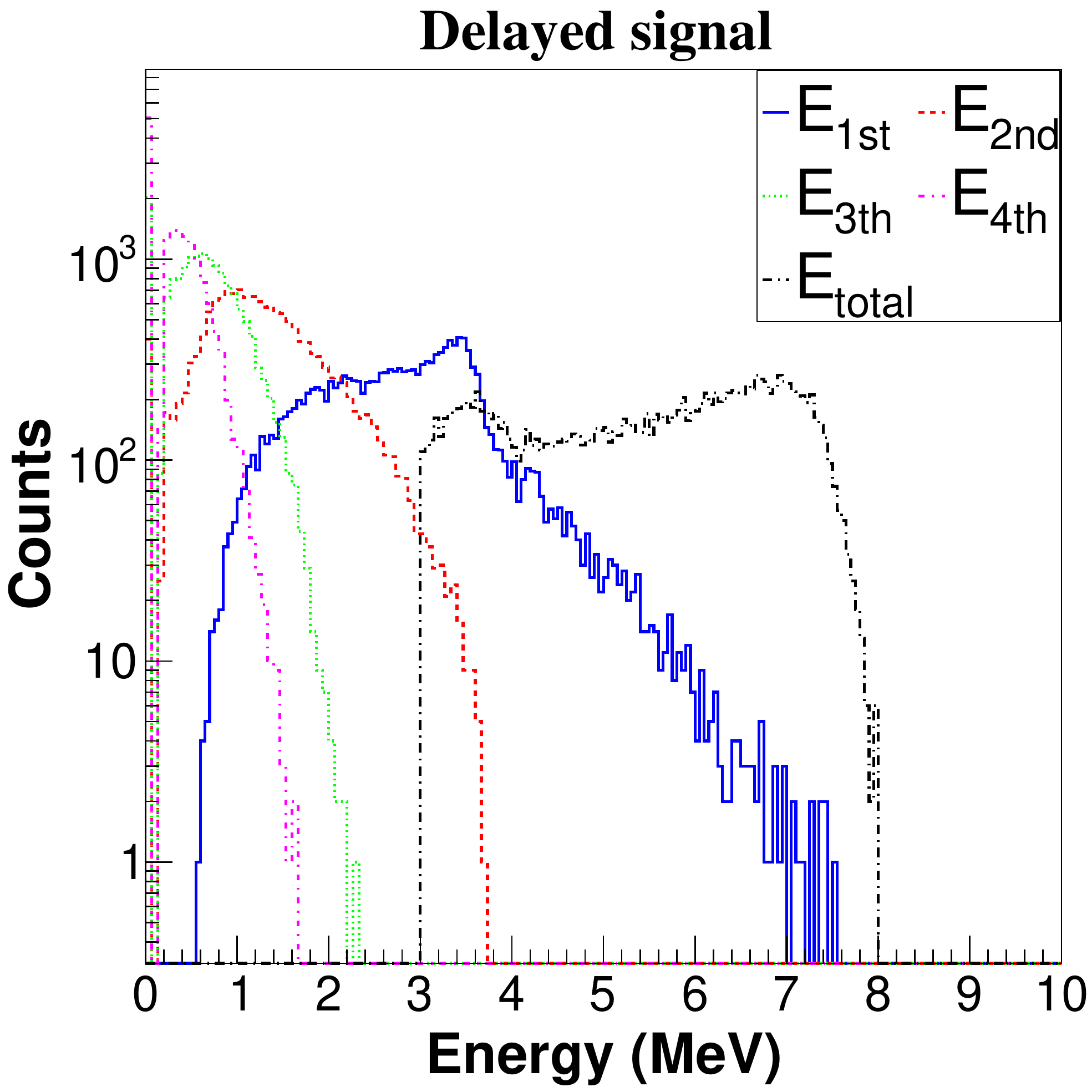}
        \label{fig:fig12b}
    }
    \caption{ (a) The deposited energy distribution of single fast neutron background events which satisfy 2.5 MeV $<$ $E_{total}$ $\leq$ 8 (prompt), 3 MeV $<$ $E_{total}$ $\leq$ 8 (delayed), and 4 $<$ $\Delta$T $<$ 200 cuts.  }
    \label{fig:fig12}
\end{figure} 

We then apply further cuts ($E_{1st}$, $E_{2nd}$, $E_{3th}$, $E_{4th}$ and $N_{hit}$ ) to these passed events (shown in Fig. $\ref{fig:fig12}$) to estimate the rejection efficiency of each selection criterian. The effect of each selection criterion on single fast neutron rejection efficiency is presented in Table $\ref{table:3}$. The $E_{2nd}$ $\leq$ 520 keV requirement for the prompt signal is found to be very effective in distinguishing IBD events from single fast neutron background events. A total of 96$\%$ rejection efficiency is estimated when all the selection cuts are applied.

For multiple fast neutron backgrounds, we simulate neutrons over a wider energy range (1keV-50MeV)\footnote{It is seen from Fig. $\ref{fig:fig11}$ that neutron energies less than 1 MeV fall further into the delayed energy interval of 3-8 MeV.}. Two independent delayed signals from different neutron simulations are combined and one of the delayed signals is considered as prompt signal. Again, the events passing the $E_{Total}$ (both prompt and delayed) cut are taken into account. Additional cuts ($E_{1st}$, $E_{2nd}$, $E_{3th}$, $E_{4th}$ and $N_{hit}$ ) are then applied to these generated data to estimate the rejection efficiency of each selection criterion. It is found that the selection condition of $E_{2nd}$ and $N_{hit}$ is very effective to discriminate between the prompt events and the delayed events. The results are presented in Table $\ref{table:3}$.

\begin{figure}[!htb]
    \centering
    
        \includegraphics[width=0.48\linewidth]{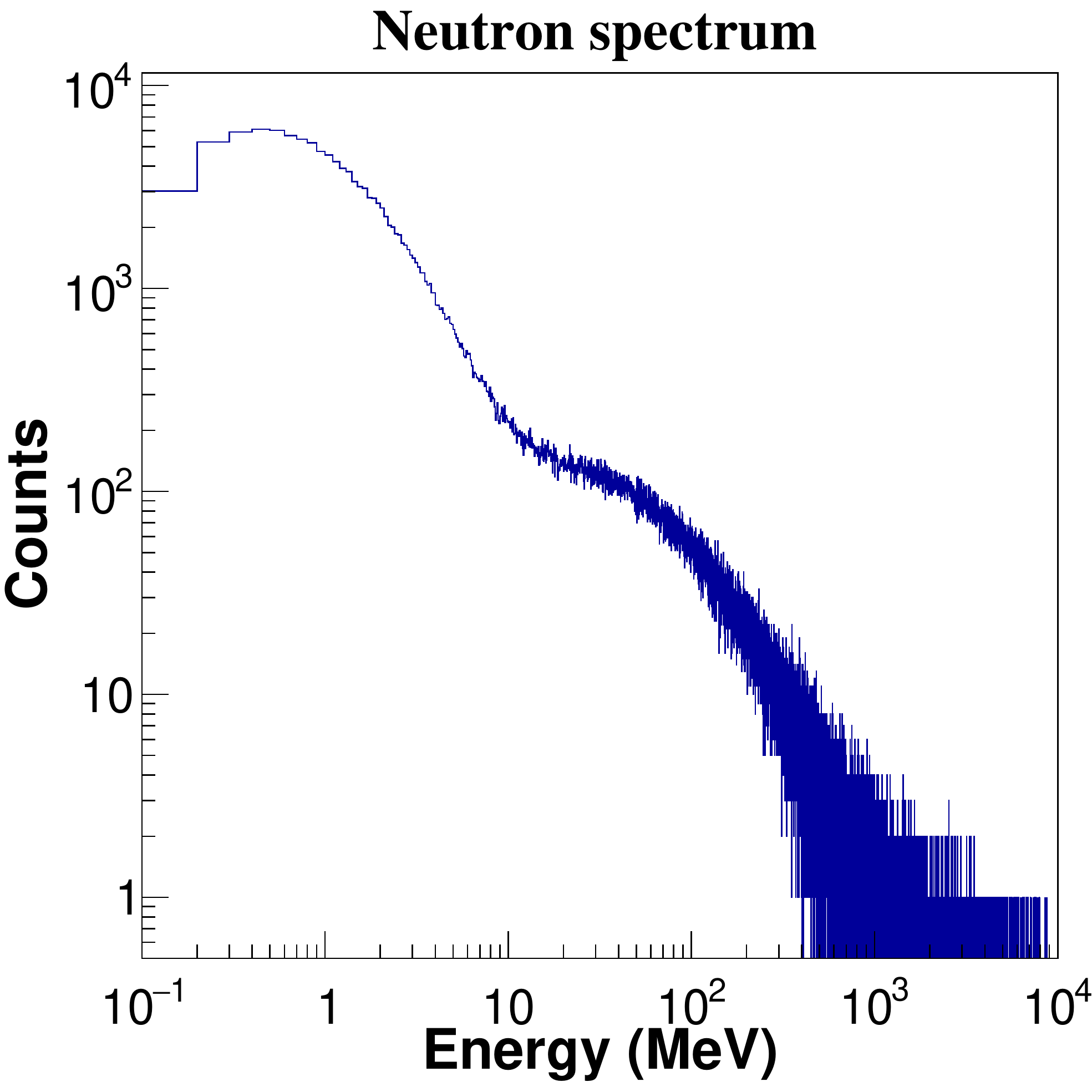}
        \caption{ Initial neutron energy distribution used in the simulation. The initial neutron energy is generated according to the neutron energy spectrum measured on the ground by Gordon \cite{Gordon}. }
        \label{fig:fig13}
\end{figure}

Finally, since the proposed detector operates aboveground, we use a real neutron spectrum measured on the ground by Gordon \cite{Gordon} as an input in the simulation (Fig. $\ref{fig:fig13}$). We then calculate single fast neutrons and multi-neutrons rejection efficiency as calculated above. The results are presented in Table $\ref{table:3}$.

\clearpage
\section{Conclusion}

We have presented a plastic scintillator-based segmented antineutrino detector to monitor power and fissile content of Akkuyu NPP. The proposed detector prefers to use hexagonal plastic scintillator bars in contrast to conventional parallelepiped. Although hexagonal bars require a custom design, combining hexagonal bars into a honeycomb fashion provides more compactness and thus lessens the number of optical readout channels required for a given detector volume. The segmentation structure of the detector allows discriminating the true IBD event from the background by forming a unique pattern of each passing track. An antineutrino signal analysis has been developed using the selection technique based upon both the topology and relative timing of the prompt and delayed signal. A list of selection conditions is established to precisely tag the antineutrino events. A detection efficiency of 10$\%$ is estimated from Monte Carlo simulation with the selection cuts applied. Even with this low efficiency, it has been shown that a few hundred antineutrinos can be detected per day at a reasonable distance from the reactor (20-30m). In addition, it has been shown that backgrounds resulting from single fast neutrons and multiple fast neutrons can be rejected with an efficiency of 96$\%$ and 98$\%$, respectively.

\clearpage




\bibliographystyle{elsarticle-num}

\bibliography{sample}

\end{document}